\documentclass{jltp}
\usepackage{amssymb}

\usepackage[dvips]{graphicx}

\begin{document}

\title{$\Pi $ \`{a} la node: disordered $d$-wave\\
superconductors in two dimensions for the random masses}
\author{P. J. Hirschfeld$^1$ and W.A. Atkinson$^2$}
\date{\today }
\maketitle

\begin{abstract}
\vskip.2cm

We review work on the problem of disorder in the 2D $d$-wave superconducting
state, and show that the  symmetries of the normal state and the
disorder distribution are vital for  understanding 
the low-energy behavior. Most previous
theoretical results for the density of states (DOS) are reconciled by a
combination of exact numerical solutions of the Bogoliubov-de Gennes
equations and weak localization calculations, which suggest that a novel
diffusive mode with momentum $(\pi ,\pi )$ is responsible for a divergence
of the DOS in the globally particle-hole symmetric case. 
We note briefly that the simple
problem of a disordered tight-binding band of normal electrons displays some
similar effects, which have been overlooked in the literature.\ Finally, in
the physically realistic case of binary alloy disorder, no particle-hole
symmetry, and an order parameter which is supressed around each impurity
site, a power law with nonuniversal exponent is predicted. \ 

\end{abstract}

\address{$^1$Physics Dept., Univ. Florida, PO Box 118440,
Gainesville FL 32611}

\address{$^2$Physics Dept., Southern Illinois Univ., Carbondale, IL
62901-4401 }


\section{Introduction}

Shortly after the early work of BCS, it was recognized that
while nonmagnetic impurities would not affect thermodynamic properties of a
conventional superconductor,\cite{anderson} they might do so in hypothetical 
$p$-wave pairing systems\cite{balian}. \ With the observation of a high density of
low-energy excitations in superconducting heavy fermion systems, such states
were considered as candidates for the ground state of compounds like UBe$%
_{13}$ and UPt$_{3}$. \ Gorkov and Kalugin\cite{gorkovkalugin} and,
independently, Ueda and Rice\cite{uedarice} pointed out that linear nodal
regions of the order parameter on the 3D Fermi surface would lead to a
finite residual density of states\ at zero energy, $\rho (0).$ This
model was analogous to that used by Abrikosov and Gorkov\cite{AG} in their
discussion of gapless superconductivity caused by magnetic scatterers,
namely a self-consistent second-order perturbation theory in the impurity
potential $U$ averaged over disorder. \ Pethick and Pines\cite{pethickpines}
noted that weak scattering models of this type  
were insufficient to describe transport
properties measured on the heavy fermion systems, and proposed that large 
effective impurity scattering strengths, due to strong correlations, 
required a partial 
resummation of perturbation theory.  Hirschfeld et al.\cite%
{hirschfeldtmatrix} and Schmitt-Rink et al.\cite{schmittrink} subsequently
calculated the self-consistent $t$-matrix for impurities with phase shifts
close to $\pi /2$, and found a\ pole in the residual density of states close
to the Fermi level which broadened into a disorder-induced ``plateau'' at
low-energies, yielding metallic behavior of the nodal quasiparticles at low
temperatures in the superconducting state. Stamp\cite{stamp} then
investigated the local density of states around a single impurity within
this model.

With the proposal of $d$-wave symmetry in the hole-doped cuprate
superconductors in the early 1990s\cite{scalapino}, the self-consistent $t$%
-matrix approximation (SCTMA) was used to provide a qualitative explanation
of the effects of disorder on the $d$-wave state which provided important
supporting evidence. \ For example, observations of a crossover of the London
penetration depth from a $T$ to a $T^{2}$ behavior with a small
concentration of Zn impurities were not initially understood to be a natural
consequence of $d$ symmetry, because it was expected that in \ such
experiments $T_{c}$ would be strongly supressed in a $d$-wave system. \ In
the unitarity limit, however, strong modifications of low-energy properties
occur over an energy scale $\gamma $ (``impurity bandwidth'') which can be
much larger than the normal state impurity scattering rate $\Gamma $ which
controls $T_{c}$\cite{goldenfeld}.\ \ Analyses of transport properties also
provided evidence for near-unitarity scattering, and helped rule out
candidate extended $s$-wave states\cite{borkowski}.

\begin{figure}[th]
\begin{center}
\includegraphics[width=0.6\columnwidth]{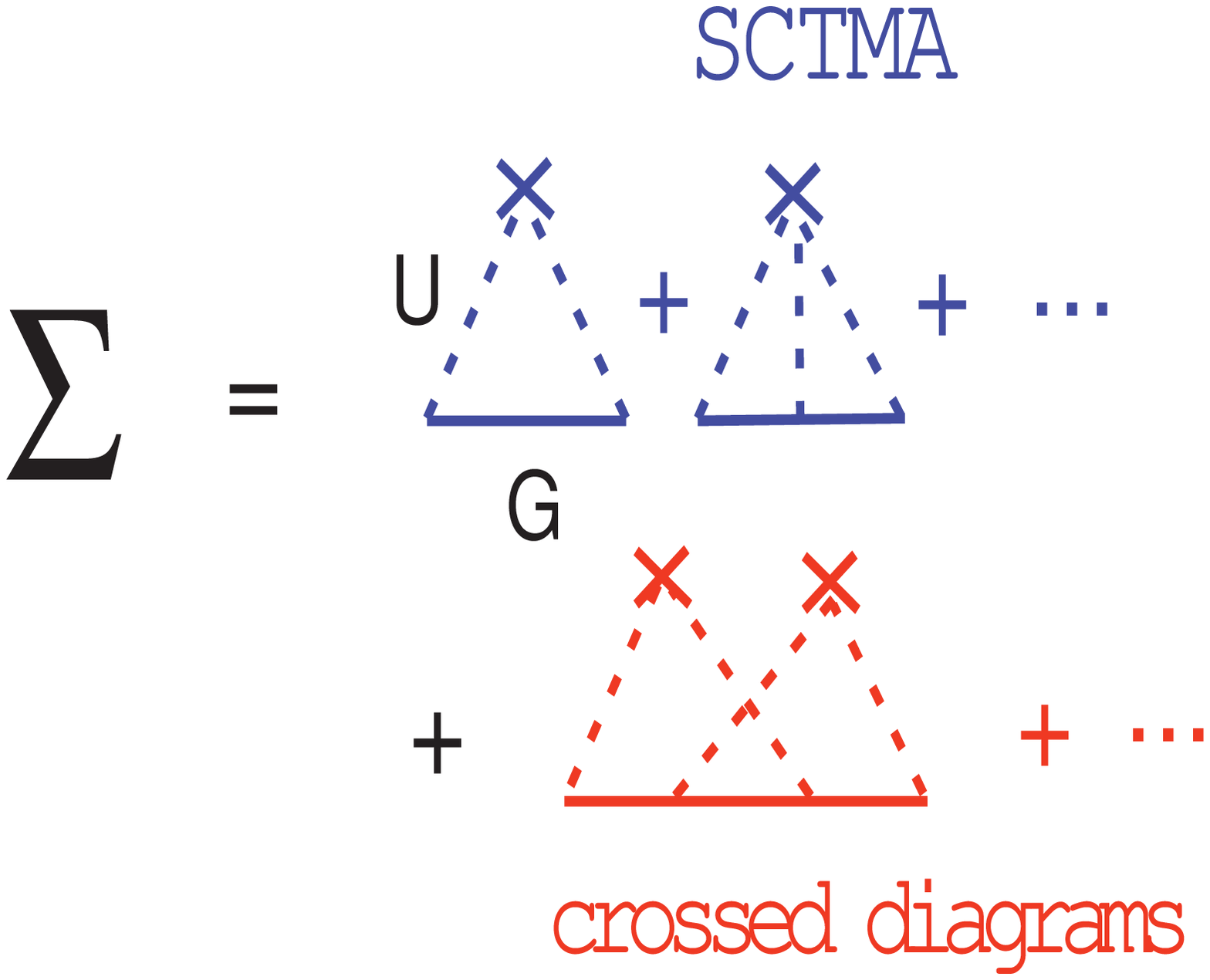} \label{fig:one}
\end{center}
\caption{Self-energy after disorder averaging.}
\end{figure}
This relatively successful phenomenology was cast into doubt in a work by
Nersesyan, Tsvelik and Wenger\cite{tsvelik}, who pointed out that the
utility of the $t$-matrix approach was based on the ability to neglect
multiple-impurity scattering processes (``crossed diagrams''), which are of
higher order in $1/k_{F}\ell $ or the density of impurity sites $n_{i}$ in
three dimensions. \ These diagrams have an additional $\log E$ singularity
in 2D coming from the line nodes of the $d_{x^{2}-y^{2}}$ gap, however,
which leads to a breakdown of single-site perturbation theory and the need
to resum at least a second class of crossed diagrams. \ This important paper
led to a series of attempts to solve the problem nonperturbatively,
following in most cases standard methods which had produced reliable results
in the case of the disordered normal metal. \ By 2000 several such solutions
had appeared in the literature\cite%
{tsvelik,ziegler1,fisher,zirnbauer,pepinlee}, but if anything the situation
was more confused, since the various papers had each claimed to calculate 
\textit{exactly} the density of states of a disordered $d$-wave
superconductor, but had arrived at dramatically different results. \ In
particular, the claimed residual densities of states at zero energy were
zero \cite{tsvelik,zirnbauer,fisher}, finite\cite{ziegler1}, and infinite%
\cite{pepinlee}, respectively. \ 

In this paper, we review efforts we began at that time to reconcile the
various available theoretical results. 
We emphasize that our interest was primarily to
understand real $d$-wave superconductors with strong nonmagnetic disorder,
rather than classify models and search for universal physics. We also sought
to understand which, if any, of the nonperturbative solutions was relevant
to the ``realistic'' model and to experiments on cuprate superconductors.  
Most of this work, involving
both numerical solutions of the Bogoliubov-de Gennes equations and
analytical weak localization calculations, has been published previously.%
\cite{atkinson1,atkinson2,atkinson3} \ \ In these papers it was shown that 
most of the various nonperturbative solutions are correct, and that subtle
differences in symmetries of the normal state band and disorder models led
to the conflicting results. In particular, the divergent density of states
found in ``globally particle-hole symmetric models'' (we use this language
to describe models with normal state bands manifesting particle-hole
symmetry on all energy scales, not just near the Fermi surface) was found to
arise due to a qualitatively new diffusive mode with momentum $(\pi ,\pi )$.
It is interesting to note that this sensitivity to ``details'' of disorder
does not exist in the theory of disorder in normal metals, but may be
thought of as a critical behavior of the $d$-wave state due to the linearly
vanishing density of states.

The general principle that random systems of different symmetry should
display qualitatively different low-energy behavior in the DOS has
its origins in random matrix theory and was extended
to $d$-wave superconductors by Senthil and Fisher\cite{fisher}
and Bocquet et al.\cite{zirnbauer} 
While this concept is very powerful, it has not always been possible to
calculate the asymptotic low-energy DOS in a controlled way using field
theoretical methods. \ This is particularly true for strong binary alloy
disorder, which such methods have  found difficult to treat.\cite{mudry}
 Very recently, however, some progress in this
direction has been reported.\cite{newfieldtheories} Perhaps more
importantly, there are no methods currently available other than numerical
ones to treat off-diagonal (order parameter) disorder \textit{correlated}
with diagonal impurity locations. \ Since the $d$-wave order parameter is
supressed essentially to zero around strong scattering centers, this would
appear to be an important effect, and we indeed find within BdG
calculations that the case where
order parameter response is self-consistently included in the numerical
solutions to the Bogoliubov-de Gennes equations is qualitatively different
from the case where the order parameter is assumed to be homogeneous. We
argue that the
physics of this correlated order response is relatively unexplored, and may
lead to important changes in low energy transport\cite{atkinsoncond},
possibly even in three dimensions. 

 In the final part of this paper, we
present some new results for a lattice model of a normal metal where global
nesting symmetry is found to play a role similar to the $d$-wave case,
and show that previous analyses have come to some erroneous conclusions.

\section{\protect\bigskip Review of nonperturbative approaches}

In their original papers, Nersesyan et al.\cite{tsvelik} considered models
in which the scattering between nodes was artificially restricted, and
mapped the problem onto a disordered chain which they then solved by
bosonization, leading to predictions of \textit{vanishing }density of states
at the Fermi level, $\rho (E)\sim E^{\alpha }$, with $\alpha $ either a
universal constant 1/7 if quasiparticles were allowed to scatter across the
Fermi surface, or a coupling constant-dependent result if only forward
scattering within a node were allowed. \ Ziegler et al.\cite{ziegler1} then
pointed out that a $d$-wave analog of the so-called Lloyd model in normal
metals was exactly soluble\ for $\rho (E)$ over the whole frequency range,
and yielded a constant density of states at $E\longrightarrow 0.$ To
complete the proof, however, they were forced to assume a Lorenztian
disorder distribution, as well as a rather strange normal state electronic
dispersion, $\varepsilon _{k}=-t^{\prime \prime }(\cos 2k_{x}+\cos 2k_{y})$%
(3rd nearest neighbor hopping). Nersesyan et al.\cite{NTWcomment}
criticized this work as nongeneric due to the use of the Lorentzian
distribution, which has long-range tails with divergent moments, but Ziegler
et al.\cite{ziegler2} were able to prove rigorous lower bounds to $\rho(0)$ 
for other
distributions. \ Shortly thereafter, Altland and Zirnbauer\cite%
{altlandzirnbauer97} considered extending the Wigner-Dyson symmetry
classification of disordered systems to superconductors (i.e. to systems
with a Bogoliubov-de Gennes-type particle hole symmetry) and
superconductor-normal interfaces. \ They argued that the number of possible
symmetry classes (and, by implication, universal low-energy behavior) for
disordered systems was restricted to the 10 Cartan Lie algebras, four of
which apply to systems with BdG symmetry.\cite{mudrytable} 

 It may be
worthwhile to review the general idea at this stage. \ The  classification
of disordered systems is based on the observation that some special
symmetries obeyed by a random Hamiltonian for every realization of the
disorder endow the statistical ensemble of Hamiltonians with the structure
of a Lie algebra. The special symmetries are, in the context of tight-binding
Hamiltonians describing the hopping of an electron in a random environment:
time-reversal symmetry (TR), spin-rotation symmetry
(SR),
particle-hole symmetry, and sublattice symmetry.  For a bipartite lattice,
the sublattice or ``chiral" symmetry alluded to corresponds to the
statement that the Hamiltonian changes sign when all sites on
sublattice A are multiplied by -1 (see below).  Clearly an on-site
potential breaks this symmetry, which is therefore very special.
 That these four symmetries are
exhaustive from the point of view of group theory follows from the
classification by Cartan of all possible finite-dimensional symmetric
spaces. \ Combination of the scaling approach to localization and the Cartan
classification of symmetric spaces then leads to the constructions of
non-linear sigma models with the Cartan symmetric space as a target space to
describe the physics of localization in the presence or absence of these
special symmetries.

 Senthil and
Fisher\cite{fisher} were the first to investigate how different symmetries
affected the DOS of a disordered $d$-wave system.  They first 
calculated the quantum corrections to the DOS and
conductivity in the ``metallic state'' defined by the SCTMA ``plateau'', an
analogy originally suggested by Lee\cite{palee}, and showed that 
the DOS was supressed as $\delta
\rho (E)$ $\simeq$ $-\rho _{0}(v_{\Delta }/8v_{F})\log \gamma /E$ for $E\lesssim
\gamma $, where $v_{F}$ is the Fermi velocity, $\rho _{0}$ is the plateau
DOS, and $v_{\Delta }$ is the gap velocity near the nodes, hinting at a
vanishing DOS at $E=0$. In addition, they gave a heuristic argument for an
effective 1D model whose low energy properties should correspond to the
4-node $d$-wave system, finding $\delta \rho (E)\sim E$ below a strong
localization scale $E_{1}\sim \gamma \exp -v_{F}/v_{\Delta }$ for
a $d$-wave superconductor with TR- and SR-invariance (Cartan class CI), 
and $\rho(E)\sim E^2$ if
TR is broken (class C).     Bocquet et
al.\cite{zirnbauer} mapped a more general problem of disordered Dirac
fermions onto a supersymmetric field theory which could be solved by RG
methods under certain conditions. \ They \ found as a special case that for
disordered d-wave superconductors with both broken TR and
SR symmetry (class D), the density of states diverges
logarithmically as $E\rightarrow 0$. \ \ The class C and D results are 
irrelevant to the
current work, in which we concern ourselves exclusively with singlet
superconductors in zero external field with nonmagnetic disorder
(generically class CI), \ but illustrated the importance of symmetry for the
DOS and localization properties in the $d$-wave superconducting state. \ 

All the above results were restricted to weak scattering, in the sense of a
narrow distribution of site energies. \ Noting that in the cuprates there
was considerable evidence for localized unitarity limit scattering centers, P%
\'{e}pin and Lee\cite{pepinlee} embarked upon a new approach to the disorder
problem, considering formally the exact $t$-matrix for a given configuration
of $N$ unitary scatterers in a half-filled tight-binding \ band. \ They
predicted that the DOS was singular as $E\rightarrow 0$, and claimed that
the leading such singularity was described by scattering $N$ times from a
single impurity, leading to $\delta \rho (E)\sim 1/(E\log ^{2}\Delta /E)$
 (an integrable divergence).

In Table 1, we collect the results of several nonperturbative calculations
for disordered $d$-wave superconductors. For each reference, we indicate
with a few words which model was considered and the Cartan symmetry class of
the model in question. It will become clear from the discussion below that
the symmetry class in each case was not always obvious to the authors of the
work in question, and in some cases is still not settled. 

\bigskip 
\begin{table}[tbp]
\begin{tabular}{|l|l|l|l|}
\hline
group & class & model/method & result $\rho (E\rightarrow 0)$ \\ \hline\hline
{Nersesyan et al.\cite{tsvelik}} & AIII & Dirac fermions & $\sim |E|^{\alpha
}$ \\ 
~ &  & w/ random gauge fields, & ${\alpha <1}$ dep. on {\# nodes,} \\ 
~ &  & supersymm./bosoniz., & ~internode scattering \\ \hline
{Ziegler et al.\cite{ziegler1}} & AI & Lloyd model & $\sim $ const.+ $aE^{2}$
\\ \hline
{Senthil-Fisher\cite{fisher}} & CI & $Sp(2n)$ nonlin. $\sigma $ model, & $%
\sim |E|$ \\ 
&  & 2+$\epsilon $ expansion &  \\ 
& C &&$\sim E^2$\\
\hline
{P\'{e}pin-Lee\cite{pepinlee}} & AIII?  & Leading sing. of $t$-matrix, & $\sim
1/(|E|\log ^{2}\Delta /|E|)$ \\ 
~ &  & ${1\over 2}$-filled unitarity limit & ~ \\ \hline
{\ Bocquet et al.\cite{zirnbauer}} & D & Dirac fermions/NL$\sigma $ & $\sim
\log |E|$ \\ 
&  & nonabelian bosoniz. &  \\ \hline
{Altland et al.\cite{zirnbauer3}} & CI & Dirac fermions/NL$\sigma $... & $%
\sim |E|$ \\ 
& AIII &  & $\sim |E|^{\alpha }$ \\ 
& C &  & $\sim E^2 $\\ 
& A &  & const. \\ \hline
Fabrizio et al., & AIII? & NL$\sigma$ model & $\sim e^{-A\sqrt{\log |E|}}/|E|$ \\
{Altland\cite{newfieldtheories}}&&${1\over 2}$-filled unitarity limit  &\\
\hline
\end{tabular}%
\caption{Nonperturbative approaches to disordered $d$-wave DOS problem.}
\end{table}

\section{\protect\bigskip Numerical results with homogeneous order parameter}

Before reviewing numerical finite-size calculations, we remind the reader of
some details of the $t$-matrix approach. \ We focus on the effect of
point-like disorder, characterised by two parameters, the impurity
concentration $n_{i}$ and the potential $\hat{U}$. The latter is described
by a $\delta $-function scattering potential, $\hat{U}(\mathbf{R}-\mathbf{R}%
_{imp})=U_{0}\delta (\mathbf{R}-\mathbf{R}_{imp})\tau _{3}$, where the $\tau
_{i}$ are the Pauli matrices in particle-hole space. The impurity $t$-matrix
is given as usual by (in matrix notation) $\hat{T}=\hat{U}+\hat{U}\hat{G}_{0}%
\hat{T}$, with $\hat{U}$ defined above. \ \ The result is \ 
\begin{equation}
\hat{T}(\omega )\simeq {\frac{g_{0}-c\tau _{3}}{c^{2}-g_{0}^{2}}},
\label{eq:tmat}
\end{equation}%
where $c=U_{0}^{-1}-g_{3}$, and $g_{0}$ and $g_{3}$ are the components of
the momentum integrated Green function, $g_{\alpha }\equiv (1/2)\sum_{k}%
\mathrm{Tr\;\tau }_{\alpha }{\hat{G}}(\mathbf{k},E)$. \ \ For the single
impurity problem, Eq. (\ref{eq:tmat}) is exact, and ${\hat{G}=\hat{G}}^{0}$.
\ The poles of this expression correspond to impurity resonances at position 
$\Omega _{0}$ with finite width due to coupling to the $d$-wave continuum,
unless $\Omega _{0}=0$. \ From Eq. (\ref{eq:tmat}) it is clear that, since $%
g_{0}\sim E$ up to log corrections, the resonance is \textit{not} located
exactly at the Fermi level for infinitely strong potential $U_{0}=\infty $,
unless $g_{3}=0$. \ This occurs only for a \textit{globally }particle-hole
symmetric electronic spectrum, e.g a single tight-binding band at half-filling. \ The
disorder-averaged self-energy is now defined in the limit of a density $%
n_{i} $ of independent impurities to be $\hat{\Sigma}(\mathbf{k},E)\equiv
n_{i}\hat{T}_{\mathbf{kk}}(E)$, and determined self-consistently with the
averaged $\hat{G}$ via the Dyson equation, 
\[
\hat{G}^{-1}=E-\xi _{k}\tau _{3}-\Delta _{k}\tau _{1}-\hat{\Sigma}(\mathbf{k}%
,E)\equiv \tilde{E}-\tilde{\xi}_{k}\tau _{3}-\tilde{\Delta}_{k}\tau _{1}. 
\]%
In the SCTMA the impurity band is characterised by the scattering rate $%
\gamma =-\mathrm{Im}\Sigma (E\rightarrow 0)$, that also approximately
determines the energy at which $\rho (E)$ crosses over from linear ($%
|E|>\gamma $) to constant ($|E|<\gamma $). \ It is this constant behavior
which Nersesyan et al\cite{tsvelik} have shown is not generally applicable
in 2D.

The true low-energy behavior of the 2D system can be calculated within mean
field theory by solving the BdG equations. \ In this paper we present
results for a tight-binding lattice with $N=1600$ sites and up to 50
disorder configurations. In matrix form, the mean-field Hamiltonian is 
\begin{equation}
\mathcal{H}=\sum_{ij}\Phi _{i}^{\dagger }\left[ 
\begin{array}{cc}
t_{ij} & \Delta _{ij} \\ 
\Delta _{ij}^{\dagger } & -t_{ij}^{\ast }%
\end{array}%
\right] \Phi _{j}  \label{eq:ham}
\end{equation}%
with $\Phi _{i}^{\dagger }=(c_{i\uparrow }^{\dagger },c_{i\downarrow })$.
The subscripts $i$ and $j$ refer to site indices, and $t_{ij}=-t\delta
_{\langle i,j\rangle }+(U_{i}-\mu )\delta _{i,j}$ with $\delta _{\langle
i,j\rangle }=1$ for nearest neighbour sites, and 0 otherwise. All energies
in this work are measured in units of $t$, and the lattice constant is $a=1$%
. The bond order-parameter is $\Delta _{ij}=-V\langle c_{j\downarrow
}c_{i\uparrow }\rangle $ with $V$ the nearest neighbour pairing interaction.
The pure $d$-wave superconducting state occurs in the disorder-free limit
and is related to the bond order parameters by $\Delta _{ij}=\frac{1}{2}%
\Delta _{0}[(-1)^{x_{ij}}-(-1)^{y_{ij}}]$ with $(x_{ij},y_{ij})$ connecting
sites $i$ and $j$, and where $\Delta _{0}$ is the homogeneous $d$-wave
amplitude. The eigenstates of the finite-sized system are found using
standard LAPACK routines to diagonalise Eq.~\ref{eq:ham}, and the
quasiparticle DOS defined by $\rho (\omega )=N^{-1}\sum_{n}\delta (\omega
-E_{n})$ is then evaluated using a binning procedure. Spatial fluctuations
arise naturally when one solves $\Delta _{ij}$ self-consistently in the
presence of a disorder potential. \ For now, we simply hold $|\Delta
_{ij}|=\Delta _{0}$ fixed. \ 
\begin{figure}[h]
\begin{tabular}{cc}
\includegraphics[width=.53\columnwidth]{atkinsonfig3.eps} & %
\includegraphics[width=.47\columnwidth]{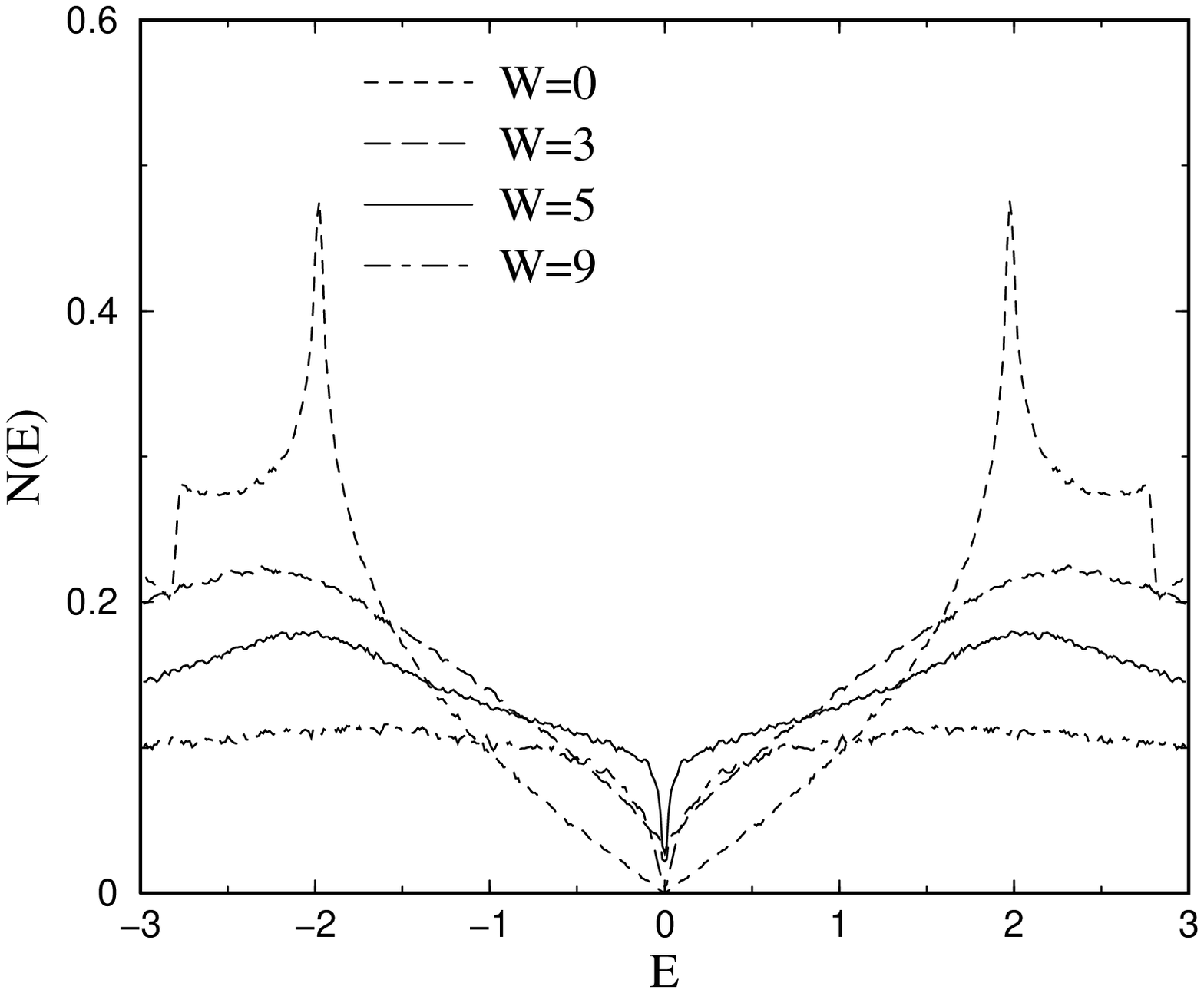} \label{fig:two} 
\end{tabular}%
\caption{a) DOS in binary alloy model near the unitary limit for $\Delta
_{0}=2 $. Main figure: $\protect\mu =0$, $n_i=0.05$. Inset: $\protect\mu =0.2
$ with $U_{0}^{-1}=0.001$ (solid), 0.01 (dotted), 0.02 (dashed).
Dashed-dotted line: SCTMA for $U_{0}^{-1}=0.001$ b) Box disorder.
Parameters: $\protect\mu =1.2$, $\Delta _{0}=2.0$.  All energy units are in 
terms of the bare hopping $t$.}
\end{figure}
We focus primarily on the binary alloy distribution, where one\ assumes $%
U_{i}=U_{0}$ $\ $on $n_{i}$ of the sites, 0 elsewhere, and then averages
over configurations. \ In this case one may compare directly with the SCTMA.
\ In Figure 2a, we show that the SCTMA result agrees quite well with the
exact result over most of the energy range.  Very close to the Fermi level,
the DOS appears
to go to zero, except for the one case of very large $U$ shown.  This scale is 
actually determined in  Figure 2a  by the
discreteness of the level spacing, consistent with the fact that the scale $%
E_{1}$ over which such supressions are expected in the infinite 
system\cite{fisher} is
exponentially small in the so-called ``Dirac cone anisotropy'' factor $%
v_{F}/v_{\Delta }\simeq 2t/\Delta_0$. \ \ There are small deviations in the curves shown
between SCTMA and the exact BdG solution, but they are confined to an energy 
$E\lesssim \gamma $; we postpone discussion of these features for the
moment. \ Note the divergence of the DOS at $E\rightarrow 0$\cite%
{atkinson2,ting} occurs \textit{only }when the chemical potential is zero
and the impurity scattering strength is infinite. \ In the language of the
SCTMA, $\mu =0$ \textit{and} $c=0$ are necessary conditions. \ Deviations
from either of these symmetries splits the central peak and forces $\rho
(E)\rightarrow 0$ at $E=0$. \ The \textit{form }of the divergence is
difficult to deduce directly from the finite-size numerical results,
although a claim has been made that the P\'{e}pin-Lee prediction provides a
reasonable fit\cite{ting}. \ It is worth noting that one can also split the
peak as described with the addition of a next-nearest neighbor hopping to
the Hamiltonian, thereby destroying the global particle-hole symmetry of the
band even at $\mu =0.$ On the other hand, if one performs the same numerical
``experiments'' described above with the Ziegler et al. 3rd nearest neighbor
dispersion, a \textit{constant }DOS is always obtained\cite{atkinson2}, in
agreement with Ref. \cite{ziegler1}.

Random site energy distributions, which we can also study with numerical
methods by distributing $U_{i}$ randomly (Gaussian, Lorentzian, ``box''
disorder), never give rise to the divergences seen in the binary alloy
models. \ The density of states is again supressed to zero at sufficiently
low energy, but the width of this supression is unobservably small for
realistic Dirac cone anisotropies and disorder values (see Fig 2b).\cite%
{atkinson2} \ We have studied artificial cases with $v_{F}/v_{\Delta
}\approx 1$, and in fact do find results consistent with a linear variation
of $\delta \rho (E)$ for strong disorder\cite{fisher,atkinson2}, but it is
clear this result has no relevance for the real cuprates. \ It is
furthermore obvious that the usual models with random site distributions
mimic weak impurity scattering, as discussed in Ref. \cite{atkinson2}, and
therefore miss these features of strong binary alloy disorder. \ An attempt
has been made to treat unitarity scattering 
by 
conjecturing a duality symmetry connecting the perturbation expansions
in the weak and
strong scattering regimes\cite{mudry}. \ The prediction of this work is that
a divergence should result whenever $c=0$ for the single impurity problem. 
Note that this point of view implies that one should always be able to
``fine-tune'' the scattering potential to create a divergence in the
disordered system, even away from the symmetric band case. While the idea is
appealing on physical grounds, our numerical results do not support this
conjecture (see Fig. 2a, and discussion in the next section).  It appears
that the ansatz of Ref. \cite{mudry} is 
insensitive to nesting due to the neglect of 
crossing diagrams, which spoil the duality at low energies. 

\bigskip

\section{Weak localization approach}

\begin{figure}[h]
\begin{center}
\includegraphics[width=0.9\columnwidth]{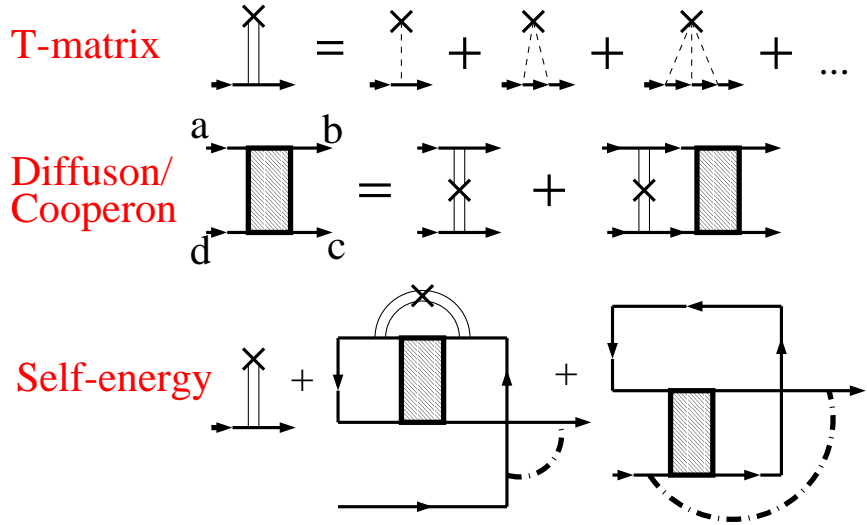} \label{fig:three}
\end{center}
\caption{Diagrams necessary for calculation of quantum correction to DOS.
Dashed line is impurity interaction $\hat U$, and solid line is $d$-wave
Nambu propagator $\hat G$.}
\end{figure}
The above numerical analysis is striking and does provide an explanation at a
superficial level for the differences in the various nonperturbative
approaches to the disordered $d$-wave problem\cite{atkinson2}. \ On the
other hand, despite careful scaling calculations, any rigorous conclusions
are necessarily limited by the finite size of the numerical experiments. In
addition, the numerics do not provide any insight into the origin of the
novel physics reflected in the divergence seen at half-filling and strong
scattering, or other aspects of the results. \ For these reasons we have
investigated analytically the leading quantum corrections to the density of
states in the $d$-wave superconductor, following Senthil and Fisher\cite%
{fisher} in the weak scattering case, and extending the calculation to
arbitrary scattering potential\cite{atkinson3}.

We first calculate the particle hole (diffuson $\mathcal{D}_{\mathbf{q}}$)
and particle-particle (Cooperon $\mathcal{C}_{\mathbf{q}}$) propagators,
keeping diagrams to formal leading order in $1/\sigma $, where $\sigma $ is
the dimensionless conductance\cite{atkinson3}.
These diagrams are shown in Figure 3, along
with the self-energy processes to the same order which determine the DOS. \
The usual diffusion mode associated with particle conservation is recovered
at low energies,

\begin{equation}
\mathcal{D}_{\mathbf{q}}(E,E^{\prime })=\mathcal{C}_{\mathbf{q}}(E,E^{\prime
})=\frac{\gamma ^{2}}{\pi \rho _{0}}\frac{1}{\mathcal{D}_{0}q^{2}-i(E-E^{%
\prime })},  \label{diffuson}
\end{equation}%
\qquad \qquad \newline
$\ $where $\mathcal{D}_{0}={\frac{v_{F}^{2}}{2\gamma}} \log \frac{%
v_{F}v_{\Delta }}{\gamma }$ is the quasiparticle diffusion constant. \ The
gapless mode in the Cooperon channel drives the leading singularity at low
energies, yielding the Sentil-Fisher result for $\delta \rho (E)$ given
above,\textit{\ provided }no further gapless diffusive modes exist. \ This
is precisely the case if the Green's function has, for a given impurity
configuration, an additional symmetry relating its values on the two
sublattices of the underlying crystal. \ We consider first the 
half-filled tight binding model on a square lattice with unitarity scattering case
treated by P\'{e}pin-Lee, which diplays the nesting (sublattice
or ``chiral") symmetry 
$\tau _{2}\hat{G}_{\mathbf{k}%
}\tau _{2}=\hat{G}_{\mathbf{k+Q}}$.
Note again that the symmetry condition is only
fulfilled if the potential is infinite, i.e. it effectively removes a site
entirely from the lattice. \ It is easy to check that any second nearest
neighbor hopping, model for the gap going beyond the simple nearest neighbor
bond pairing model we have adopted, or any finite chemical potential will also
break the symmetry. \ Any finite on-site potential $U_{0}$ does the same,
regardless of whether or not the single-impurity resonance is fine-tuned to
lie at the Fermi level, in contrast to the conjecture in Ref. \cite{mudry}. 
\begin{figure}[h]
\begin{center}
\includegraphics[width=0.9\columnwidth]{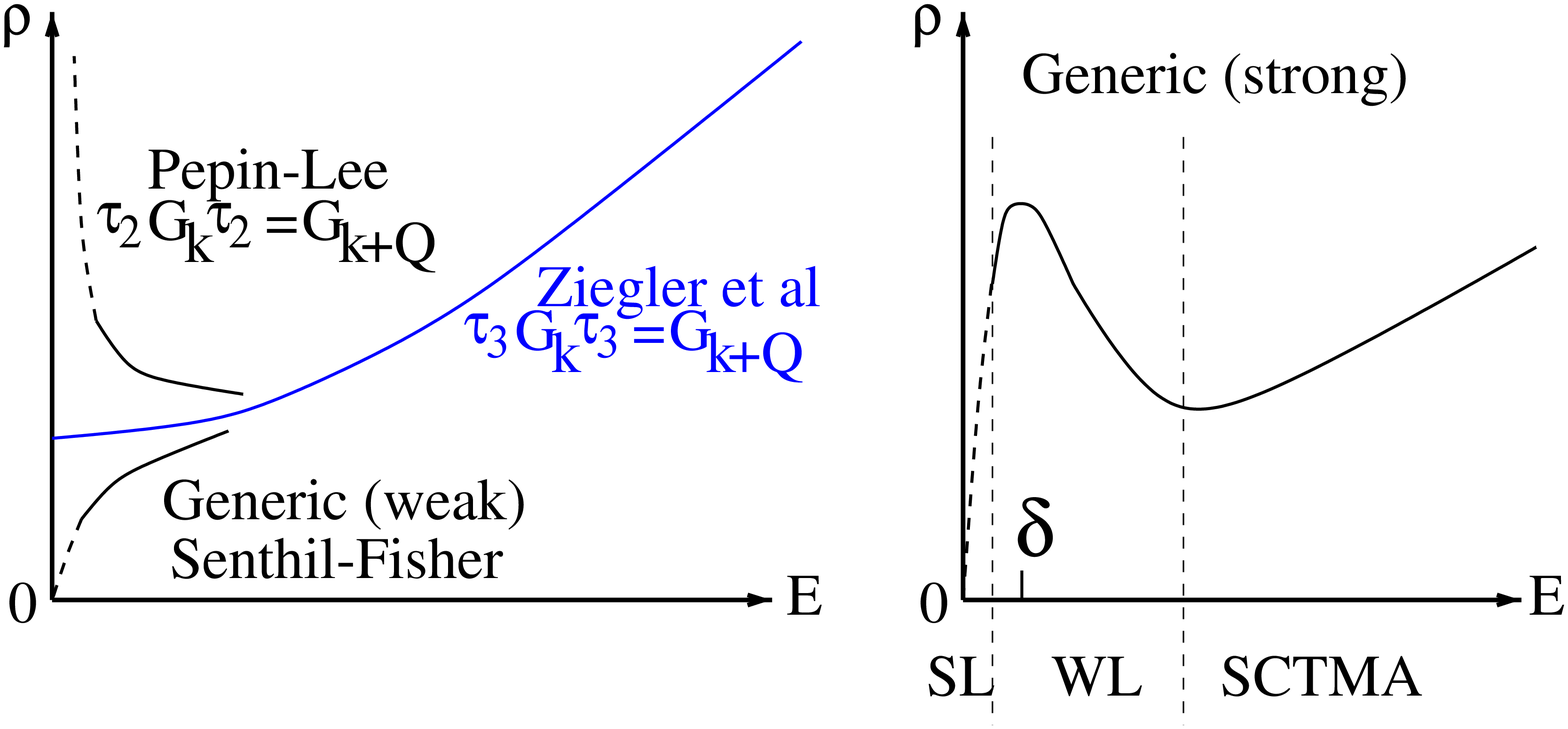} \label{fig:four}
\end{center}
\caption{Schematic dependence of low-energy DOS on symmetry class. \ SL and
WL indicate strong and weak localization regimes, respectively.}
\end{figure}

The extra symmetry induces a new massless diffusive mode in the Cooperon
channel at $\mathbf{q}=(\pi ,\pi )$ leading to the divergence in $\rho (E)$.
\ How this occurs may be seen by evaluating the DOS slightly away from the
symmetric point,%
\begin{equation}
\delta \rho (E)\simeq \frac{1}{2\pi \mathcal{D}_{0}\log \frac{v_{F}v_{\Delta
}}{\gamma }}\left( -\log \frac{\gamma }{2E}+5\log \frac{\gamma }{\sqrt{%
4E^{2}+\delta ^{2}}}\right) ,  \label{pidos}
\end{equation}%
where $\delta $ is now the mass of the new $\pi $-mode. \ If the mass is
finite, the Senthil-Fisher \textit{supression }of the DOS is recovered at
energies below roughly $\delta /2$. \ On the other hand, if the mass
vanishes, or if $\delta <E<\gamma $, the DOS will increase with decreasing
energy due to the sign of the second factor in 
Eq. (\ref{pidos})\cite{atkinson3}. How the
mass varies with deviations from the symmetric point determines the
intermediate-energy behavior of the DOS; for example, for infinite potential
and a tight-binding band, the mass is given by $\delta \simeq \mu ^{2}/\log
(v_{F}v_{\Delta }$/$\gamma )$. It worth noting that if one tries to
calculate this quantity in the nodal approximation, one can be easily
misled; away from half-filling for infinite potential, its value is
determined entirely by the Fermi surface curvature near the nodes. Finally,
the Ziegler et al model can also be seen in this picture to be qualitatively
different because the Hamiltonian contains no terms which mix
the two sublattices; formally the Green's function obeys the symmetry $\tau
_{3}\hat{G}_{\mathbf{k}}\tau _{3}=\hat{G}_{\mathbf{k+Q}}$. \ In this case,
all singular contributions to $\rho (E)$ can be shown to cancel to leading
order, consistent with a constant value at $E=0$\cite{atkinson3}.  Altland and Zirnbauer%
\cite{zirnbauer3} have pointed out that this model is equivalent in its
localization properties to the symmetry class of a normal metal (class AI);
it is nevertheless a true superconductor with a Meissner response. \ Since
its electronic dispersion is very unphysical, this model, while soluble, has
no predictive power for the cuprates.\ \ \ In Figure 4 we show the schematic
behavior of the DOS for the various cases we have considered. Note that the
functional form of the DOS at asymptotically low energies (below $E_{1}$) is
not determined by this method, as indicated by the dashed portions of the
curves.

Before leaving the subject of nesting symmetries in the homogeneous case, we
ask what a symmetry analysis might have told us \textit{a priori} about the
low energy behavior of the DOS. Although the Hamiltonian of a bond-paired $d$%
-wave superconductor with half-filled band and unitarity limit scatterers
obviously displays a Bogoliubov-de Gennes symmetry, it should not belong to
the Cartan symmetry classes C,CI,D,DIII usually associated with
superconductors. \ According to the usual definitions, only CI corresponds
to the unbroken time-reversal and spin-rotation symmetry we consider here;
the density of states for this class corresponds to the Fisher-Senthil\cite%
{fisher} result $\rho (E)\sim E$, as discussed above. \ The obvious
conclusion is that the additional nesting symmetry combines with the BdG
symmetry operation to move the system into a new Cartan class; this has been
recently determined to be the unitary chiral metal class AIII.\cite%
{newfieldtheories} However, it is puzzling that both the single-node model
originally studied by Nersesyan et al.\cite{tsvelik} and the half-filled
unitary scattering case both correspond to class AIII (see Table 1); 
the former model yields a power law DOS $\rho (E)\sim |E|^{\alpha }$,
whereas the latter gives a divergence at small $E$.

The answer to this particular puzzle is that the  
Cartan classification is a purely algebraic (group theoretical) one
which must sometimes be refined by allowing topological
considerations, i.e., by deciding whether or not a physical model should be
approximated in the context of the scaling theory of localization by a
non-linear sigma model with the addition of possible topological terms. In
this way models belonging to the same Cartan class can nevertheless display
strikingly \textit{different} localization properties. The most famous
example thereof is the integer quantum Hall effect that naively belongs to
class A but whose localization properties are described by a non-linear
sigma model supplemented by a topological term. 
The model of Nersesyan et al. is also endowed with such a topological 
term,\cite{zirnbauer3}  leading to the deviation from 
the naive AIII prediction.  In addition, the low energy behavior depends
on the strength of the disorder, and for sufficiently strong disorder 
gives a constant density of 
states.\cite{gurarie}

Thus knowledge of the Cartan symmetry class is 
\textit{not} a priori sufficient to fix the low-energy behavior; a model
calculation is required. In the background is a more general set of
questions raised in Refs.\cite{cartanquestions} concerning whether a given
disordered electron system ``automatically'' belongs to one of Cartan's
classes, or whether it requires some fine tuning of the disorder
distribution. \ We do not address these issues here.

   It is also
obvious that there is a further discrepancy evident in Table 1, in
that the {\it same} model is claimed to have two different asymptotic
behaviors of the DOS as $E\rightarrow 0$ in Refs. \cite{pepinlee}
and \cite{newfieldtheories}.  
This is why we have indicated the symmetry class of the model
in both cases as AIII?.
At present it is unclear
if numerical approaches can distinguish between these two predictions,
but one of them must be incorrect.

\section{Inhomogeneous order parameter}

Thus far we have considered numerical and analytical calculations with
constant order parameter only. \ \ In self-consistent calculations with
strong scatterers, $\Delta _{ij}$ vanishes (or nearly vanishes) along bonds
connected to the impurity site, and rises to its asymptotic value over a few
lattice constants. The naive expectation is that the pockets of normal metal
which form around the impurities should enhance $\rho (0)$, but numerical
studies show that the superconducting gap tends to reopen\cite{atkinson1}
relative to the SCTMA. \ In Figure 5, we exhibit this tendency for a
realistic parameter set (i.e. no special nesting symmetries). 
\vskip 1cm
\begin{figure}[h]
\begin{center}
\includegraphics[width=.95\columnwidth ]{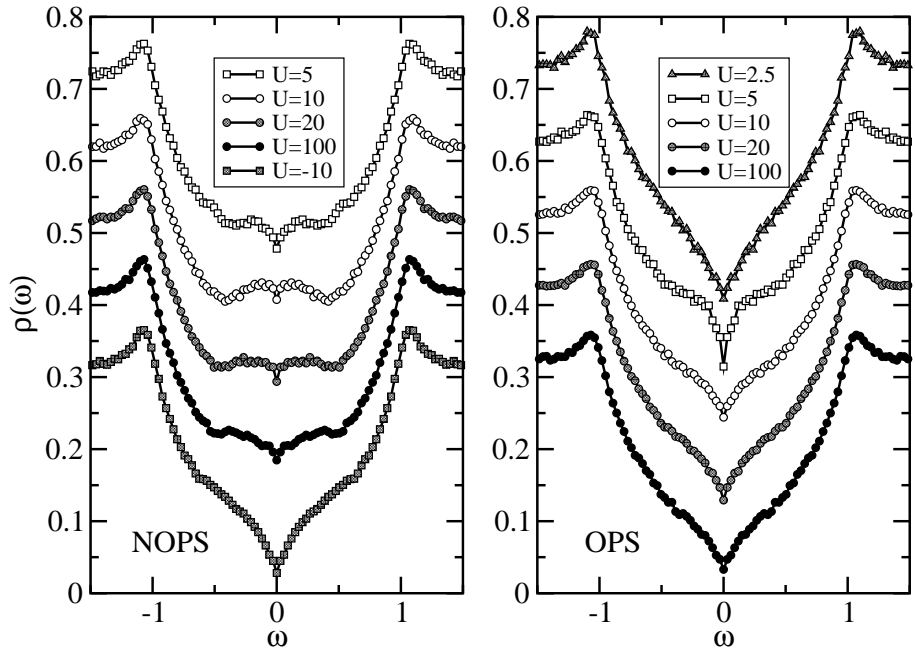} \label{fig:five}
\end{center}
\caption{Comparison of density of states with homogeneous (A) and
self-consistently determined (B) order parameter. Parameters: $n_{i}=0.06$, $%
\protect\mu =1.2t$, $\Delta _{0}=0.8$.}
\end{figure}
\vskip .2cm
We note that, close to unitarity (defined by $\Omega _{0}=0$), the
homogeneous gap case is essentially constant at the Fermi level except for
the single point at zero energy supressed due primarily to finite size
effects. By contrast, even close to unitarity the DOS exhibits a cusp-like
behavior. Away from unitarity, the disorder-induced supression occurs on a
scale which we have argued\cite{atkinson1} scales with $\Omega _{0}$. Note
this quantity is renormalized by the inhomogeneity\cite{atkinson4} in a
nonuniversal way. \ The behavior of the low-energy DOS in these calculations
is quite different from the non-self-consistently determined order parameter
case. In Figure 6, we show that over a wide range at low energies, the DOS
can be approximated by a power law, whose exponent is \textit{nonuniversal},
in that it depends on the impurity potential, concentration, etc. Of course,
we can make no statements about the asymptotic low-energy form below some
exponentially small strong localization scale. It is conceivable, even
likely, that the lowest energy behavior is again given by the universal
class CI (Senthil-Fisher) behavior. However, is at least interesting to ask
whether off-diagonal (in particle-hole space) disorder \textit{correlated} \
with bare (diagonal) disorder does not correspond to a different Cartan
symmetry class, or whether it is amenable to this classification scheme at
all. No field theoretical treatments of this problem are available to our
knowledge. Again, we note that the asymptotic low energy behavior is, in any
case, unlikely to be observed in experiment since the Dirac cone anisotropy $%
v_{F}/v_{\Delta }$ is always significantly greater than one.
\vskip .8cm
\begin{figure}[h]
\begin{center}
\includegraphics[width=.95\columnwidth]{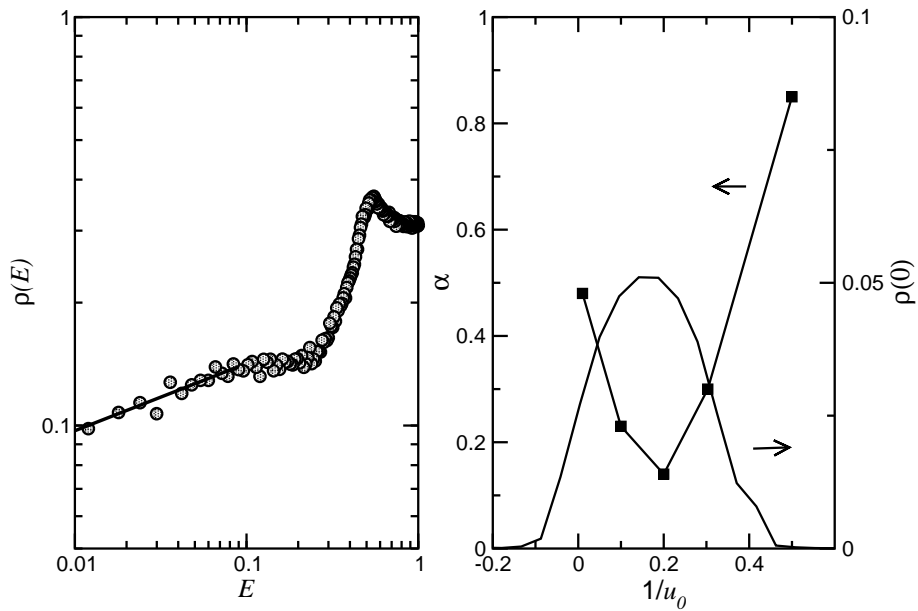} \label{fig:six}
\end{center}
\caption{Low-$E$ power law DOS for self-consistently determined order
parameter. Left panel: solid line is fit to power law $\protect\rho(E)\sim
|E|^\protect\alpha$ for $n_i=0.04$ and $U_0=5$, close to unitarity limit $%
\Omega_0=0$. Right panel: power law exponent $\protect\alpha $ and SCTMA $%
\protect\rho(0)$ vs. impurity potential $U_{0}$ for $n_i=0.02$. }
\end{figure}

The\ physical origin of the disorder-induced supression of the DOS relative
to SCTMA is not completely clear. Despite the correlation with the
single-impurity resonance energy, it appears to be a genuine many-impurity
interference effect. We speculate that it may be formally related to the
physics of the Coulomb gap in interacting disordered metals,\cite%
{altshuleraronov,ES} which can be captured by the Hartree-Fock approximation
for the Coulomb interaction.\cite{MacDonald} Here, in analogous fashion, we
treat the pairing interaction at the mean-field (BdG) level. Further work
along these lines is in progress.

We conclude this section by noting that, while the generic case of the
inhomogeneous order parameter is quite different from the homogeneous case,
the nesting symmetries continue to play a role. For example, the $\rho(E)$
divergence at zero energy in the $\tau_2$ nesting case is preserved when one
calculates the order parameter self-consistently, although its width is
renormalized\cite{atkinson2}.

\section{Half-filled normal metal}

Since the nesting symmetries discussed in Section 4 occur also for normal
metals in special cases, it is interesting to return to these simpler
problems to see what one can learn. A 2D simple tight-binding band of
electrons displays already in the pure case a weak logarithmic van Hove
singularity at $E=0$. Naively, one would expect generic disorder to mix the $%
k$-states near the Fermi level and cut off the weak nesting singularity,
leading to finite DOS at half filling in the presence of any disorder.
However, Nakhmedov et al.\cite{Nakhmedov} recently considered this problem
in a weak localization framework, and came to the conclusion that the DOS
was infinite. Such a conclusion is not without precedent in normal metals.
Dyson studied 1D chains with chiral symmetry, i.e. systems where
neither the pure Hamiltonian nor the disorder potential (e.g., random
hopping) breaks the sublattice symmetry $\tau _{2}\hat{G}_{\mathbf{k}}\tau
_{2}=\hat{G}_{\mathbf{k+Q}}$. Gade and Wegner\cite{Gade} discussed a 2D
example of a disordered nonlinear sigma model obeying the same symmetry. In
these cases the DOS at zero energy was indeed shown to be infinite.
However, in the more generic case studied by Nakhmedov et al., the on-site
potential scattering explicitly breaks the sublattice symmetry; Gruzberg et
al.\cite{Gruzberg} therefore criticized the claim of infinite DOS, insisting
the van Hove singularity should not affect the generic finite DOS one
expects in a normal metal. They redid the weak localization calculation,
obtaining a finite result. 
\begin{figure}[h]
\begin{center}
\includegraphics[width=.8\columnwidth]{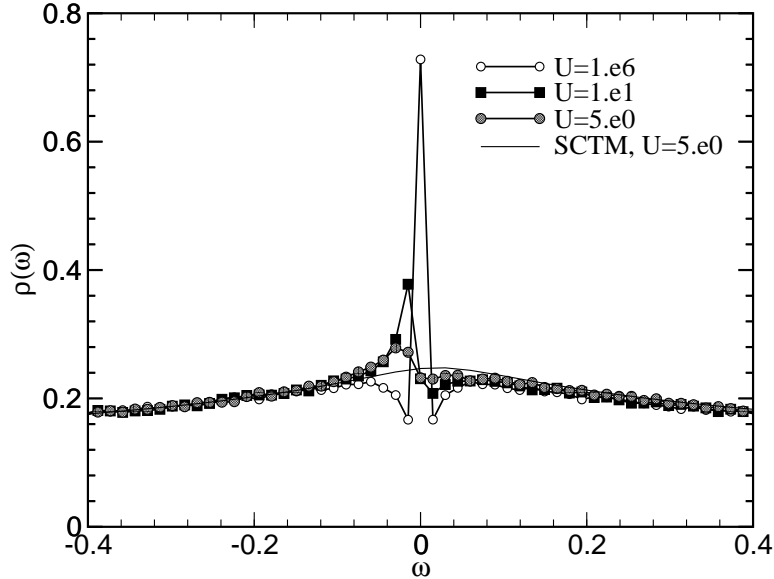} \label{fig:seven}
\end{center}
\caption{ Density of states of simple tight-binding model on square lattice
at half-filling, for
binary alloy potential $U_{0}=5,\ldots ,10^{6}$. Concentration of impurities is
$n_i=0.1$. }
\end{figure}

It is of course easy to apply the numerical scheme developed for the $d$%
-wave superconductor to the half-filled tight-binding band. Our results are
shown in Fig. 7. For finite impurity potential strength, the DOS exhibits no
divergence at $E=0$, supporting the claim of Gruzberg et al. On the other
hand, for $U_{0}\rightarrow \infty $, a narrow peak does appear. While with
numerical methods we are unable to distinguish between a finite peak at zero
energy and a true divergence, it is striking that this behavior appears to
occur only when the global sublattice symmetry $\tau _{2}\hat{G}_{\mathbf{k}%
}\tau _{2}=\hat{G}_{\mathbf{k+Q}}$ applies ($\hat{G}$ is now diagonal) for a
given disorder configuration. The infinite potential simply removes a site
from the lattice and therefore does not break the sublattice symmetry.
Formally, this system is in Cartan symmetry class BDI, whereas the finite
potential case belongs to class AI. This distinction has been thoroughly
discussed and understood in in disordered quantum wires\cite{mudrywire}, and
on this basis we indeed anticipate that the numerical result in Figure 6
indicates a divergence for the case of infinite 
potential.\cite{mudrycomment}  Hints of the effects of this nesting symmetry
were observed already in models with random hopping.\cite{eilmes}

\section{Conclusions}

In this paper we have tried to review a variety of approaches to the problem
of disorder in the $d$-wave state at low energies. While we succeeded, using
a combination of numerical and analytical methods, in understanding the
sensitivity of this state to different symmetries of the band and of the
disorder realization, most of the issues involved were of a purely
theoretical nature. This is due to several factors: first, the strong
localization scale is exponentially small in the Dirac cone anisotropy
(expected to be of order 10 in the cuprates), so within the calculations
with homogeneous order parameter the SCTMA proves to be adequate, modulo
some minor bumps at intermediate energies, if one works sufficiently far
from half-filling. Secondly, the nesting symmetries considered do not apply
in real systems, as they are always broken by finite impurity potential,
nonzero chemical potential, and $t^{\prime}$ hopping. Next, the systems in
question are weakly three dimensional, and the mass in the $\pi$ mode will
always saturate at a 3D crossover scale we estimate to be $t_\perp^2/\gamma$%
, where $t_\perp$ is the interlayer hopping. On this basis, we have
suggested  that if the physics of low-dimensional correlated scattering is
ever to be observed experimentally, the best chances are in underdoped,
intrinsically anisotropic materials where the interlayer coupling is
weakest. There are some preliminary results that such a downturn in the DOS
indeed occurs in highly underdoped LSCO\cite{lin}, but the situation is far
from clear at the present writing. All other experimental results on
systematically disordered cuprates of which we are aware
(performed almost exclusively
on systems near optimal doping) are consistent with a finite
disorder-induced DOS.

The final effect we have discussed is the profound influence of order
parameter inhomogeneity induced by and correlated with disorder in the $d$%
-wave superconductor. This phenomenon renders the problem considerably less
amenable to analytical treatment, and certainly warrants further study. Weak
localization effects are the primary cause of differences between the exact
BdG calculations with homogeneous order parameter and the SCTMA, and it is
clear that they will be irrelevant in a truly 3D system, where we know the
crossed diagrams are negligible. On the other hand, we have no real evidence
that the effect of order parameter inhomogeneity weakens significantly in
higher dimensions, since there is no well-controlled ``mean field theory of
disorder" which includes it explicitly\cite{hh}. It is interesting to ask
whether these effects can strongly influence transport, since they
fundamentally change the nature of scattering even at the single-impurity
level, by adding a scattering channel with a completely different (Andreev)
symmetry. Preliminary calculations of the conductivity in these models
indicate that this is indeed the case\cite{atkinsoncond}.

\vskip .5cm

\textit{Dedication. \ }This paper is gratefully and affectionately dedicated
by PJH to Peter W\"{o}lfle on the occasion of his 60th birthday. \ \vskip %
.2cm \textit{Acknowledgements.} The ideas in this paper were developed with
the other authors of Refs. \cite{atkinson1,atkinson2,atkinson3} (A.H.
MacDonald, K. Ziegler, A. Yashenkin, D. Kveschenko, and I. Gornyi) over two
years, and we have benefitted greatly from further discussions with A.
Altland, I.
Gruzberg, C. P\'{e}pin, and A. Shytov. PJH is particularly grateful to C.
Mudry for his patience in explaining the Cartan classification. Work was
partially supported by NSF grant DMR-9974396.


\begin{thebibliography}{99}
\bibitem{resdos} Note that the finite size numerical procedure used cannot
determine whether or not the DOS at zero energy is actually zero, or small
and finite, since energy levels are discrete and the DOS is smoothed by a
binning procedure. \ All results are however consistent with a zero residual
DOS unless otherwise stated.
\end{thebibliography}
\end{document}